\documentclass[10pt,conference]{IEEEtran}
\makeatletter
\renewcommand{\@IEEEsectpunct}{~}
\makeatother

\usepackage{cite}

\ifCLASSINFOpdf
  \usepackage[pdftex]{graphicx}
\else
  \usepackage[dvips]{graphicx}
\fi
\usepackage{amsmath}

\usepackage{multicol, multirow, tabularx}
\usepackage[table]{xcolor}
\usepackage{csquotes}
\usepackage[normalem]{ulem}
\usepackage{tcolorbox}
\usepackage{caption}
\usepackage{algorithm, algpseudocode}
\usepackage[hyphenbreaks]{breakurl}

\usepackage{url}

\usepackage{csquotes}
\usepackage[T1]{fontenc}

\begin{document}
%

\title{SecretBench: A Dataset of Software Secrets}

\author{\IEEEauthorblockN{Setu Kumar Basak\IEEEauthorrefmark{1},
Lorenzo Neil\IEEEauthorrefmark{2}, Bradley Reaves\IEEEauthorrefmark{3} and
Laurie Williams\IEEEauthorrefmark{4}}
\IEEEauthorblockA{North Carolina State University, USA\\
Email: \IEEEauthorrefmark{1}sbasak4@ncsu.edu,
\IEEEauthorrefmark{2}lcneil@ncsu.edu,
\IEEEauthorrefmark{3}bgreaves@ncsu.edu,
\IEEEauthorrefmark{4}lawilli3@ncsu.edu}}

\maketitle

\begin{abstract}
According to GitGuardian's monitoring of public GitHub repositories, the exposure of secrets (API keys and other credentials) increased two-fold in 2021 compared to 2020, totaling more than six million secrets. However, no benchmark dataset is publicly available for researchers and tool developers to evaluate secret detection tools that produce many false positive warnings. \textit{The goal of our paper is to aid researchers and tool developers in evaluating and improving secret detection tools by curating a benchmark dataset of secrets through a systematic collection of secrets from open-source repositories.} We present a labeled dataset of source codes containing 97,479 secrets (of which 15,084 are true secrets) of various secret types extracted from 818 public GitHub repositories. The dataset covers 49 programming languages and 311 file types.
\end{abstract}


%
\IEEEpeerreviewmaketitle

\section{Introduction} \label{Introduction}
GitGuardian reported in March 2022 that the number of secrets leaked on GitHub repositories doubled in 2021 compared to 2020, totaling more than six million secrets~\cite{GitGuardian}. Often, a software program uses third-party services, including payment systems, location services, and social media integration. Software developers need secrets (API keys, access tokens, and private keys) to authenticate these third-party services as part of system integration. However, developers may expose these secrets in plain text in the version control systems (VCS) or the application packages~\cite{Meli2019HowBC,mobile-app-leak}. Although the problem with checked-in secrets is well known, the secret leakage incidents continued. On September 2022, Uber confirmed an organization-wide cybersecurity breach because of having hard-coded secrets in a PowerShell script~\cite{uber-breach}. The attackers got the administrator access and compromised Uber's AWS, GCP, Google Drive, and Slack workspaces.

To avoid exposing secrets in VCS, several open-source and proprietary secret detection tools~\cite{list-secret-detection-tools}, such as TruffleHog~\cite{trufflehog} and Microsoft CredScan~\cite{cred-scan}, are available. However, these tools have been shown to produce false positive warnings~\cite{rahman2022secret}. In previous studies~\cite{saha,sinha}, researchers have worked on reducing false positives. However, their curated datasets are not large and varied and are unavailable for future research and evaluation purposes. In addition, developers face challenges in choosing one tool out of many, and no publicly-available dataset is available for comparing the effectiveness of the tools. 

\textit{The goal of our paper is to aid researchers and tool developers in evaluating and improving secret detection tools by curating a benchmark dataset of secrets through a systematic collection of secrets from open-source repositories.}

We present SecretBench, a labeled dataset of source codes consisting of 97,479 secrets extracted from 818 public GitHub repositories using two secret detection tools. We manually inspected each secret and labeled 15,084 secrets as true secrets. The dataset encompasses 49 programming languages and 311 file types. The dataset is hosted in Google BigQuery~\cite{google-big-query} and Cloud Storage~\cite{google-cloud-storage} and designed to be amenable to expansion by the community. Our dataset will aid in expediting the research to evaluate and improve secret detection tools.

\section{Data Extraction} \label{DataExtraction}
We provide our eight-step process for data collection of SecretBench as follows:

\textbf{Step 1: Open Source Software Repository Platform Selection:} We choose GitHub~\cite{github} to select candidate repositories containing secrets for our study. GitHub is the most popular platform for hosting open-source software development projects~\cite{about-github}. As of December 2022, GitHub has over 94 million developers and more than 330 million repositories~\cite{about-github}, including at least 36 million public repositories~\cite{github-public-repo-count}.

\textbf{Step 2: Build Regular Expression (Regex) Pattern Set:} We build a regex pattern set for different types of secrets to identify the candidate repositories containing secrets for our study. For example, the regex pattern for a Slack token is ``\texttt{(xoxb|xoxp|xapp|xoxa|xoxr)\-[0-9]{10,13}\-[a-zA-Z0-9\-]*}''. TruffleHog~\cite{trufflehog}, a popular open-source secret-scanning tool, has a package of secret detectors~\cite{trufflehog-detector-package}. We extracted 751 regex patterns from the source code of the detector package and included those in our pattern set. In addition, we included 10 regex patterns from Meli et al.~\cite{Meli2019HowBC} to find the presence of secrets in GitHub repositories that are not present in the TruffleHog detector package. In total, we used 761 regex patterns in our pattern set, which is available online~\cite{regex-pattern-set}.

\textbf{Step 3: Identify Candidate Software Repositories:} To identify the candidate software repositories, we used the Google BigQuery Public Dataset of GitHub~\cite{google-big-query} (Dataset ID: \textit{bigquery-public-data.github\_repos}), which was released in 2016 by Google in collaboration with GitHub. The source code of over 2 billion files from more than 2.9 million open-source licensed repositories can be accessed with SQL queries~\cite{google-big-query}. We used the most recent snapshot available at the start of this project (September 20, 2022). We wrote an SQL script with all the 761 regex patterns to search for secrets in the source code files and executed the script in Google BigQuery. The SQL script took almost 22 hours to complete, as every file is checked with all the regex patterns. The returned result is a table of two columns: ``repo\_name'' and ``matches''. The ``repo\_name'' column represents the repository name, and the ``matches'' column represents the list of regex patterns matched with the specific repository. In total, we have found 2,234,618 repositories with at least one regex pattern match.


\textbf{Step 4: Apply Selection Criteria on Candidate Repositories:} As suggested by prior research~\cite{perils-mine-github}, GitHub repositories need to be curated by removing inactive, beginner, and tech-demo projects. To curate the repositories collected in Step 3, we collected fork information, contributor counts, and commit counts using the GitHub Rest API~\cite{github-rest-api}. We applied the following selection criteria to curate the collected repositories. The number in parenthesis with the criteria name indicates the number of filtered repositories after applying that specific criteria. 

\begin{itemize}
    \item \textbf{Availability (2,013,913):} The repository is available to download.
    \item \textbf{Uniqueness (1,735,864):} The repository is not a forked repository. This criteria is applied to avoid near duplicates of the same repository.
    \item \textbf{Collaboration (889,984):} The repository contributor count must be at or above the dataset median of 2. This criteria is applied to avoid personal or hobby projects.
    \item \textbf{Development History (622,719):} The repository commit count must be at or above the dataset median of 20 commits.
    \item \textbf{Recent Activity (93,958):} The repository must have at least one commit in the last one year. This criteria is applied to avoid inactive projects.
\end{itemize}


In addition, we observed some repositories with different ``repo\_name'' fields point to the same repository. For example, repositories ``Jasig/cas''
and ``apereo/cas'' 
are the same repository though having different repository names in the dataset. This duplication happened because the repository owner changed the repository name at some point, but the Google BigQuery dataset kept both names. However, GitHub stores the actual repository name of the duplicate repository. We collected the actual repository name of each repository using the GitHub Rest API and filtered the duplicate repositories. After all selection criteria, we passed 89,070 unique repositories to Step 5.

\textbf{Step 5: Find Multiset-Multicover Repositories of Regex Patterns:} 
In this step, we further select repositories so that we get a sample of multiple secrets for each secret type while minimizing the overall repository count of the dataset. 
In later steps, we manually determine if identified secrets were actually secret or not. 
However, identifying and manually labeling secrets from the 89,070 repositories remaining in Step 4 is impractical. 
Our goal of  identifying the smallest selection of repositories that altogether include a specified count of each identified secret pattern is actually an instance of the \emph{multiset-multicover} problem, so we applied the multiset-multicover algorithm described in Algorithm~\ref{alg:multiset-multicover}.
This algorithm is an extension of the Minimum Set Cover algorithm~\cite{Young2008} to select a minimal set of repositories covering all the regex patterns with a certain number of repositories for each pattern.


Before applying the multiset-multicover algorithm, we observe that 390 out of 761 regex patterns found no match in any repository. The median regex pattern matched 10 repositories, with 186 regex patterns matching 10 or more repositories. We term these patterns ``upper tail'' regex patterns. An additional 120 regex patterns matched between 1 and 9 repositories; we will refer to these as ``lower tail'' regex patterns. The median lower tail regex pattern matched 2 repositories. 

For a comprehensive dataset, we seek a balance between examples of common and uncommon secret types, so we applied the multiset-multicover algorithm in two phases. In Phase 1, we ran the multiset-multicover algorithm for the 186 upper tail regex patterns to find a set of repositories where each regex pattern matches at least 10 repositories. We identified 649 repositories among the upper tail regex patterns. For Phase 2, we ran the multiset-multicover algorithm for the 120 lower tail regex patterns to find a set of repositories where each regex pattern should match at least 2 repositories and identified 190 repositories. Then, we merged the repositories of Phase 1 with Phase 2 and removed duplicate repositories. Altogether, we identified 818 repositories for SecretBench to collect candidate secrets.

\begin{algorithm}
\caption{Multiset-Multicover Algorithm}
\label{alg:multiset-multicover}
\footnotesize
\begin{algorithmic}[1]
\Require $PatternsToCover, U$
\Require $InstanceSize, K$
\State $R_a \gets ReadAllRepos()$
\State $CoveredRepos, C_r \gets \emptyset$
\State $CoveredPatterns, C_p \gets \emptyset$
\While{$C_p \neq U$}
    \State $M \gets FindRepoWithMostUncoveredPatterns(R_a, C_p, U)$
    \State $C_p \gets C_p \cup FindMatchedPatternsForRepo(M, R_a)$
    \State $C_r \gets C_r \cup M$
\EndWhile
\State $R_{cc} \gets FindRepoCountPerPatternInInitialCover(C_r, R_a, U)$
\State $U_p \gets FindPatternsLessThanKInstance(R_{cc})$
\While{$len(U_p) \neq 0$}
    \State $M \gets FindRepoWithMostUncoveredPatterns(R_a, C_p, U_p)$
    \State $R_p \gets FindMatchedPatternsForRepo(M, R_a)$
    \State $C_p \gets C_p \cup R_p$
    \For{$e \:in \:R_p$}
        \State $R_{cc}[e] \gets R_{cc}[e] + 1$
    \EndFor
    \State $U_p \gets FindPatternsLessThanKRepoInstance(R_{cc})$
    \State $R_a \gets RemoveSelectedRepoFromList(M, R_a)$ 
    \State $C_r \gets C_r \cup M$ 
\EndWhile
\State $C_r \gets RemoveDuplicateRepos(C_r)$
\State \Return $C_r$
\end{algorithmic}
\end{algorithm}

\textbf{Step 6: Find Candidate Secrets:} We wrote a Python program to clone the repositories. We used GitPython~\cite{gitpython} to download all the branches of a repository and saved the files into a Google Cloud VM Instance~\cite{google-cloud-compute-engine} (OS: Ubuntu 18.04 LTS, RAM: 16 GB, Persistent Disk: 500 GB). Next, we ran two secret detection tools, TruffleHog~\cite{trufflehog} and Gitleaks~\cite{gitleaks}, to identify candidate secrets from the repositories. Both tools are widely used for secret detection and can identify secrets buried in the repository's history and logs. We used these tools since manually inspecting each file of a repository to find secrets is infeasible and would be error-prone. The tools provide a JSON output for each repository. The JSON output contains the candidate secrets with additional metadata such as the commit id, commit date, committer email, file path, start line, end line, start column, and end column of the file where secrets are matched. Next, we wrote another Python program to read each report generated by the tools and extract the candidate secrets along with the metadata. Altogether, we identified 97,479 candidate secrets present in different commits of 818 repositories, of which 27,336 secrets are unique.

\textbf{Step 7: Label Candidate Secrets:} The first and second authors manually inspected each candidate secret independently using the metadata collected in Step 6. A candidate secret is labeled as ``True'' if the secret is a true secret, otherwise labeled as ``False''. We observed the agreement of the labeling of secrets with a Cohen's Kappa~\cite{cohen-kappa} score of 0.86 between two raters, which indicates a ``near perfect agreement'' according to Landis and Koch's interpretation~\cite{landis-koch}. The disagreements were resolved after a discussion between the two raters. In our dataset, we identified 15,084 true secrets, of which 4,014 secrets are unique.

\textbf{Step 8: Developer Survey:} We conducted a developer survey to evaluate whether the committer of the secrets agrees with our label. First, we selected unique secrets committed between 2021 and 2022 to avoid recall bias~\cite{recall-bias} from the developers and identified 7,617 secrets. Since GitHub allows the developers to use a noreply email address (\texttt{user-name@users.noreply.github.com}) as the commit email address~\cite{github-no-reply}, we filtered those secrets and identified 2,115 secrets. Then, we selected 200 secrets (randomly selected to avoid selection bias~\cite{selection-bias}) and emailed the developers to know if they agreed with our labeling of the secret and the reason they disagreed. In the email, we provided the repository name, commit id with the commit GitHub link, file path, start line, end line, and a screenshot of the code where the secret is found. We received 56 responses, a  28.0\% response rate. Altogether, 44 (78.6\%) respondents fully agreed with our label, while 6 (10.7\%) respondents disagreed. The remaining 6 (10.7\%) respondents were not sure.

\section{Data Description} \label{DataDescription}
In this section, we provide brief details of our dataset. 

\subsection{Curated and Derived Fields:} 
We collected the metadata related to the secret such as repository name, commit id, commit date, committer email, file path, start line and end line. To further enrich the dataset, we have augmented the mined data with additional features that are computed or derived from the source code files and secrets. Example of computed and derived fields are ``file\_type'', ``is\_template'', ``in\_url'', ``entropy'', ``character\_set'' and ``has\_words''. An overview of our SecretBench dataset is presented in Table~\ref{dataset-description}.

\newcolumntype{e}{>{\hsize=0.01\hsize}X}
\newcolumntype{s}{>{\hsize=0.01\hsize}X}
\newcolumntype{b}{>{\hsize=0.90\hsize}X}
\begin{table} [!htb]
\footnotesize
\caption{Overview of the SecretBench Dataset}
\label{dataset-description}
\begin{tabularx}{\columnwidth} {|e | b | s |}
 \hline
 \multicolumn{1}{|l|}{\textbf{Field Name}} &
  \multicolumn{1}{l|}{\textbf{Description}} &
  \multicolumn{1}{l|}{\textbf{Data Type}}\\
 \hline \hline
 id & Unique identifier of the secret. & Integer\\ \hline
 secret & Candidate secret string. & String\\ \hline
 repo\_name & Name of the repository. & String\\ \hline
 domain & Domain of the repository such as GitHub. & String\\ \hline
 commit\_id & Commit hash where the secret is added. & String\\ \hline
 file\_path & File path where the secret is included. & String\\ \hline
 file\_type & Type of the file such as .py and .config. & String\\ \hline
 start\_line & Start line no. where the secret is present. & Integer\\ \hline
 end\_line & End line no. where the secret is present. & Integer\\ \hline
 start\_column & Start index of the secret in the start line. & Integer\\ \hline
 end\_column & End index of the secret in the end line. & Integer\\ \hline
 committer\_\newline email & Email address of the committer. & String\\ \hline
 commit\_date & The timestamp of the commit. & TimeStamp\\ \hline
 label & The ground truth label of the secret. & Boolean\\ \hline
 is\_template & Flag to indicate if the secret is a placeholder such as ``MY\_PASSWORD". & Boolean\\ \hline
 in\_url & Flag to indicate if the secret is part of URL such as ``http://user:pwd@site.com". & Boolean\\ \hline
entropy & Shannon entropy of the secret. & Float\\ \hline
character\_set & Characters used in the secret (NumberOnly, CharOnly, Any). & String\\ \hline
has\_words & Flag to indicate if any common English word~\cite{common-english-words} of at least length of 4 is present within the secret. & Boolean \\ \hline
length & Length of the secret. & Integer\\ \hline
is\_multiline & Flag to indicate if the secret is present in multiple lines. & Boolean\\ \hline
 category & The category of the secret. & String \\ \hline
 file\_identifier & Unique identifier of the file to check the secret from local system. & String\\ \hline
 repo\_iden\newline tifier & Unique identifier of the repository to check the secret from local system. & String\\ \hline
\end{tabularx}
\end{table}

\subsection{Data Characteristics}
Our SecretBench dataset is diverse in terms of different project characteristics. The dataset consists of 97,479 secrets in 818 repositories, and some repositories use multiple programming languages. For example, the repository ``paradite/hn-ratio''~\cite{paradite} consists of two programming languages: JavaScript and Shell. Altogether, our dataset repositories used 49 programming languages. The top 5 programming languages based on the number of repositories are Shell (459), JavaScript (414), Python (312), Java (180), and Ruby (172). The number in parenthesis denotes the number of repositories containing the specific language. In addition, our dataset consists of secrets present in 311 file types. The top 5 file types based on the number of secrets in those files are js (10,412), nix (8,623), json (8,132), txt (7,737), and xml (6,429). Besides, the top 5 file types based on the number of true secrets are txt (2,935), toml (1,985), js (1,583), html (1,337), and pem (813). The number in parenthesis denotes the number of secrets in the specific file type.

The secrets in our dataset are categorized into eight categories and presented in Table~\ref{secret-types}, sorted based on the number of true secrets. More details of our dataset is presented in our GitHub repository~\cite{secretbench-repo}.

\newcolumntype{e}{>{\hsize=0.2\hsize}X}
\newcolumntype{s}{>{\hsize=0.6\hsize}X}
\newcolumntype{a}{>{\hsize=0.2\hsize}X}
\begin{table} [!htb]
\footnotesize
\caption{The categories of secrets in SecretBench}
\label{secret-types}
\begin{tabularx}{\columnwidth} {|s | e |e|}
 \hline
 \multicolumn{1}{|c|}{\textbf{Category}} &
 \multicolumn{1}{c|}{\textbf{True Secrets}} &
 \multicolumn{1}{c|}{\textbf{Total Secrets}}\\
 \hline \hline
 Private Key & 5,789 & 8,584 \\ \hline
 API Key and Secret & 4,529 & 5,162 \\ \hline
 Authentication Key and Token & 3,569 & 5,833 \\ \hline
 Other & 524 & 66,690 \\ \hline
 Generic Secret &  334 & 439 \\ \hline
 Database and Server URL & 162 & 9,970 \\ \hline
 Password & 150 & 705 \\ \hline
 Username & 27 & 96 \\ \hline
\end{tabularx}
\end{table}

\subsection{Data Storage} 

Our dataset is stored as relation structured data in Google BigQuery (Dataset ID: \textit{dev-range-332204.secretbench.secrets}). Users can run SQL queries to access and expand the dataset. In addition, we stored the downloaded 818 repositories and the secret-containing individual source code files in Google Cloud Storage. When downloaded into the local system, the ``repo\_identifier'' and ``file\_identifier'' mentioned in Table~\ref{dataset-description} can be used to locate the repository and specific source code file related to the secret, respectively.   

Since our dataset is sensitive, Google BigQuery and Cloud Storage enable us to give access to the dataset to only selected groups, such as fellow researchers and tool developers. To get access to our dataset, researchers and tool developers need to contact us through email.

\section{Originality of SecretBench} \label{Originality}
Previous studies~\cite{sinha,saha,Meli2019HowBC} have extracted secrets from the GitHub repositories, but none made their dataset public for future research purposes. Saha et al.~\cite{saha} created a labeled dataset of 5000 secrets (700 true secrets) from 300 GitHub repositories using 32 regex patterns. With the dataset, they applied machine learning algorithms to distinguish true secrets. However, the repositories matched by regex patterns are not filtered for demo and inactive projects, and no information is provided on the files and languages covered. Sinha et al.~\cite{sinha} created a dataset of 84 GitHub repositories and identified pattern-based search and heuristics-driven filtering approaches to reduce the false positive detection of secrets. However, their dataset is small and contains only AWS credentials.

On the other hand, our dataset presented herein is large and diverse. We applied 761 regex patterns of different types of secrets and selected 818 GitHub repositories encompassing 49 programming languages. Our dataset consists of 97,479 labeled secrets, including 15,084 true secrets present in 311 different file types. We also provided different features related to the secret, such as whether the secret is a template or present in a URL. In addition, we will make our dataset available for future researchers and tool developers.

\section{Research Opportunities} \label{ResearchOpportuniteis}
To prevent exposing secrets in VCS, there are several open-source and proprietary secret detection tools~\cite{list-secret-detection-tools}. However, these tools are known to generate false positive warnings~\cite{saha,rahman2022secret}. Researchers and tool developers can identify different rules and patterns from false positive secrets to reduce false positive warnings. However, mining data from open-source and building ground-truth datasets is challenging and time-consuming. In this case, our SecretBench dataset can be used to circumvent the challenge and speed up the research and tool evaluation on reducing false positives. In addition, since several secret detection tools exist, developers face difficulty choosing one tool out of many. Future research is needed to aid developers in making informed choices about using different secret detection tools through an analysis of the effectiveness of the tools. In this case, our SecretBench dataset can act as a benchmark for comparing the effectiveness of the secret detection tools.

\textbf{Dataset Enhancement:} Our dataset can be further improved by including repositories from other VCS services such as GitLab and Bitbucket. In addition, we can add more features regarding secrets to help in secret detection automatically using machine learning algorithms. Example features include whether the secrets have parentheses (possible function call), begin with a \$ sign (possible variable), and have context words such as ``dummy'' and ``fake'' in the surrounding code of the secret. We released these additional features online~\cite{additional-features}.

\section{Ethics and Data Protection} \label{Ethics}
Since our dataset contains sensitive information such as true secrets and the committer's email addresses, we will distribute our dataset selectively.  Researchers and tool developers who want to use our dataset will sign a data protection agreement with us to avoid any unethical use. After that, we will give access to our dataset from Google BigQuery and Cloud Storage using their email addresses. In addition, at no point we did not attempt to use the secrets to verify the validity of the secrets. Instead, we labeled the secrets only by inspecting the secrets and the source code context of the secrets.

To validate our labeling, we only contacted the developers who committed the secrets. We did not reveal the identity of the developers to any managers or higher officials where they work. In addition, we are notifying every developer in our dataset to remove the secrets from their VCS.

\section{Threats to Validity} \label{Limitations}
In this section, we briefly discuss the limitations of our paper. \textbf{VCS Selection:} We did not consider other VCS services such as GitLab~\cite{gitlab} and Bitbucket~\cite{bitbucket}. In the future, we plan to expand our dataset by including repositories of other VCS services. \textbf{Manual Analysis Bias:} The labeling of the secrets in our dataset is susceptible to bias. To mitigate the bias, a second rater labeled the secrets independently, and we resolved the disagreements. \textbf{Recall Bias:} For the developer survey, though we have selected secrets that are committed in 2021 and 2022, the responses could have recall bias. We provided the developers with a screenshot of the secret-containing source code and additional metadata to mitigate the bias.

\section{Conclusion} \label{Conclusion}
We provide the SecretBench dataset consisting of 97,479 labeled secrets extracted from 818 GitHub repositories encompassing 49 programming languages and 311 file types. Our dataset will aid in evaluating and improving secret detection tools, thus preventing secret leakage in VCS and application packages. By adding new projects and features, we aim to expand our dataset. We invite the research community to join our effort to expand and enrich the dataset to create novel software secret management research opportunities.

\section*{Acknowledgement}

This work was supported by National Science Foundation 2055554 grant and the Google Cloud Research Credits program GCP19980904 award.  The authors would also like to thank the Realsearch research group for their valuable input on this paper.



\bibliographystyle{IEEEtran}
%
\bibliography{bibliography}



\end{document}